\documentclass[pra,twocolumn]{revtex4-1}
\usepackage{hyperref} 

\usepackage{graphicx} 
\usepackage{subfig}
\usepackage{color}
\usepackage{filecontents}
\usepackage{soul}
\usepackage{mathrsfs}
\usepackage{tikz,bm} 
\usepackage{circuitikz}
\usepackage{tikz}
\usepackage{physics}
\usepackage{mathtools}
\usepackage[T1]{fontenc}
\usepackage[outline]{contour}
\usepackage{relsize}
\usepackage{adjustbox}
\usepackage{graphicx}
\graphicspath{ {images/} }
\usepackage{booktabs,tabularx}
\usepackage{float}
\usepackage{svg}
\usepackage{indentfirst}
\usepackage{makecell}
\usepackage{array}
\usepackage{changepage}
\usepackage{adjustbox}
\usepackage{multirow}
\usepackage{comment}

\usepackage{hhline}
\usepackage{nicematrix}
\usepackage{xcolor}
\definecolor{Lred}{rgb}{1, 0.7882, 0.7882}
\definecolor{Lyellow}{rgb}{1, 0.9020, 0.6}
\definecolor{Lpurple}{rgb}{1, 0.6, 1}
\definecolor{Lblue}{rgb}{0.64,0.76,0.956}
\definecolor{Lgreen}{rgb}{0.6588, 1, 0.6235}
\definecolor{neonyellow}{rgb}{1, 0.94, 0.12}

\usetikzlibrary{shapes.geometric}
\usetikzlibrary{decorations.pathmorphing}
\usetikzlibrary{shapes.arrows, fadings}

\definecolor{mycolor}{rgb}{1,0.2,0.3}
\definecolor{brightgreen}{rgb}{1.0, 1.0, 1.0}
\definecolor{britishracinggreen}{rgb}{0.0, 0.26, 0.15}
\definecolor{cadmiumgreen}{rgb}{0.0, 0.42, 0.24}
\definecolor{ceruleanblue}{rgb}{0.16, 0.32, 0.75}
\definecolor{darkelectricblue}{rgb}{0.33, 0.41, 0.47}
\definecolor{darkpowderblue}{rgb}{0.0, 0.2, 0.6}
\definecolor{dt}{rgb}{1.0, 0.66, 0.07} 
\definecolor{emerald}{rgb}{0.31, 0.78, 0.47}
\definecolor{palatinatepurple}{rgb}{0.41, 0.16, 0.38}
\definecolor{pastelviolet}{rgb}{0.8, 0.6, 0.79}
\definecolor{br}{rgb}{0.5, 0.05, 0.01}
\definecolor{chosen_color}{RGB}{3, 207, 252}

\pgfdeclareverticalshading{pgf@lib@fade@north}{100bp}
{color(0bp)=(pgftransparent!0);
 color(30bp)=(pgftransparent!0);
 color(50bp)=(pgftransparent!80); 
 color(65bp)=(pgftransparent!100);
 color(100bp)=(pgftransparent!100)}%
\pgfdeclarefading{exponentialtop}{%
\pgfuseshading{pgf@lib@fade@north}%
}

\pgfdeclareverticalshading{pgf@lib@fade@north}{100bp}
{color(0bp)=(pgftransparent!100);
 color(35bp)=(pgftransparent!100);
 color(50bp)=(pgftransparent!80); 
 color(70bp)=(pgftransparent!00);
 color(100bp)=(pgftransparent!00)}%
\pgfdeclarefading{exponentialbottom}{%
  \pgfuseshading{pgf@lib@fade@north}%
}

\newcommand{\be}{\begin{equation}}
\newcommand{\ee}{\end{equation}}
\newcommand{\bea}{\begin{eqnarray}}
\newcommand{\eea}{\end{eqnarray}}
\newcommand*{\myeqref}[2][Eq.~]{%
\hyperref[{#2}]{#1(\ref*{#2})}%
}
\def\equationautorefname#1#2\null{%
Eq.#1(#2\null)%
}

\usepackage{dcolumn}
\usepackage{bm}
\usepackage{amssymb,amsmath}
\usepackage{color}
\usepackage{float}
\usepackage{tikz}
\usetikzlibrary{arrows}

\definecolor{DarkGreen}{rgb}{0,0.6,0.2}

\begin{document}
\title{Does True Randomness Exist? - Efficacy Testing IBM Quantum Computers via
Statistical Randomness}
\author{Owen Root$^{1}$}
\email{oroot@gradcenter.cuny.edu}
\author{Maria Becker$^{1}$}
\affiliation{
$^1$Department of Physics, Nebraska Wesleyan University, Lincoln, NE 68504, USA\\}
\date{\today}

\begin{abstract}
The fundamental principles of quantum mechanics, such as its probabilistic nature, allow for the theoretical ability of quantum computers to generate statistically random numbers, as opposed to classical computers which are only able to generate pseudo-random numbers. This ability of quantum computers has a variety of applications, one of which provides the basis for a method of efficacy testing Quantum Computers themselves. We introduce this testing method and utilize it to investigate the efficacy of nine IBM Quantum Computer systems. The testing method utilized four different quantum random number generator algorithms and a battery of eighteen statistical tests. Only a single quantum computer-algorithm combination was found to be statistically random, demonstrating the power of the testing method as well as indicating that further work is needed for these computers to reach their theoretical potential. 
\end{abstract}

\maketitle


\section{Introduction}
Quantum Computers (QCs) exploit the principles of quantum mechanics to enable the existence of quantum bit (qubit) based computers which possess advantages over transistor bit-based Classical Computers (CCs). These quantum mechanical properties give QCs the ability to perform certain tasks on time scales that would be impossible for CCs to accomplish in a similar time. A famous example of this is Shor's Algorithm, which lets QCs factor numbers in polynomial time, something that Classical computers can only do in exponential time \cite{shor's}. Recently, Shor's algorithm has been experimentally realized \cite{google_shor}. This particular development of QCs has drawn attention to the field for its relevance to cryptography and cybersecurity. A primary form of computer security, RSA (Rivest–Shamir–Adleman) encryption, relies on the inability of CCs to factor large numbers quickly. Shor's Algorithm could allow a QC to factor the large numbers necessary to break RSA encryption \cite{quantum_curious}. For this reason and others, many entities, both private and public, are actively researching and developing QCs. Additionally, QCs have become accessible to the general public and third-party researchers through applications such as IBM Quantum and its Quantum Educators program\cite{IBM_Quantum, quantum_educators}.

\begin{figure}[ht]
    \centering
    \parbox{0.45\textwidth}{
        \textbf{(a)}\\
        \includegraphics[scale=.29]{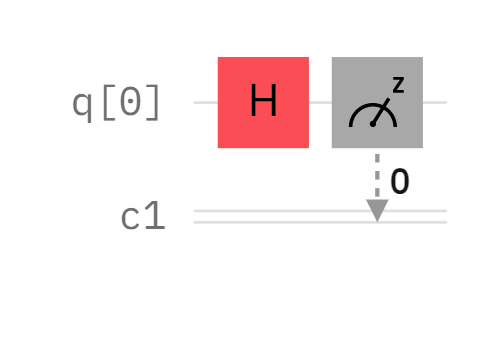}\\
    }
    \parbox{0.45\textwidth}{
        \centering
        \textbf{(b)}\\ 
        \includegraphics[scale=.29]{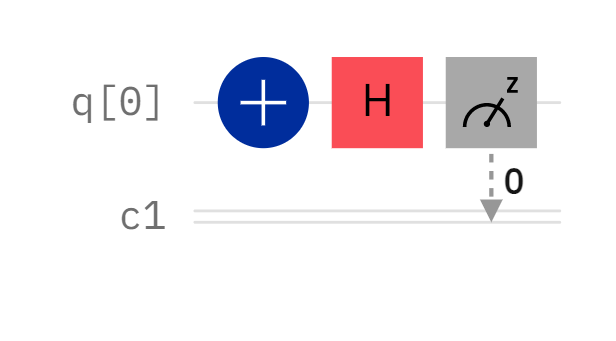}
    }
    \parbox{0.45\textwidth}{
        \centering
        \textbf{(c)}\\ 
        \includegraphics[scale=.29]{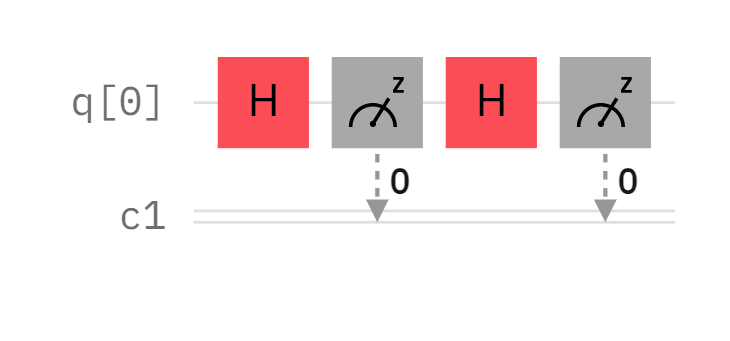}
    }
    \parbox{0.45\textwidth}{
        \centering
        \textbf{(d)}\\ 
        \includegraphics[scale=.24]{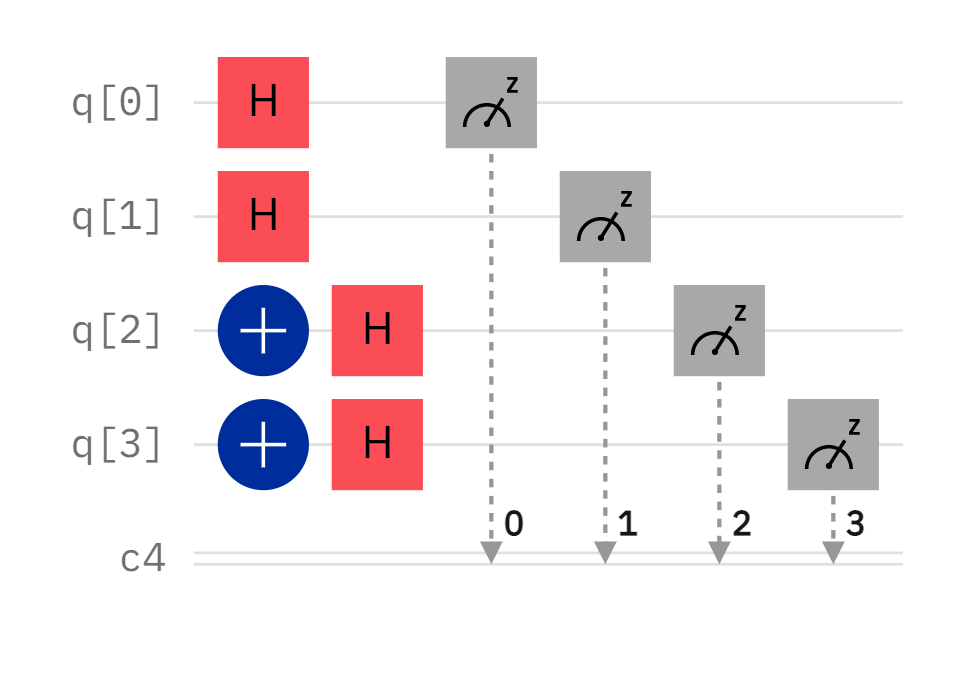}
    }
    \caption{The QRNG algorithms used in this work in standard quantum circuit notation. Blue circles with '+' are NOT gates. (a) the Basic QRNG, (b) QRNG Type 1, (c) QRNG Type 2, (d) and QRNG Type 3. Created using \cite{IBM_Quantum}.}
    \label{fig:QRNG circuits}
\end{figure}
With such widespread development of QCs, methods must be developed to assess the functional capabilities of QCs to ensure that their performance matches expectations. Here we present such a method, which operates by testing a QC's ability to act as a Quantum Random Number Generator (QRNG). 
\par
QRNGs stand in contrast to the Random Number Generators (RNGs) operated by CCs, as, due to CC's deterministic nature, RNGs are only pseudo-random.  RNGs have a significant role in classical computing, where they are used in various applications, such as cryptography, gambling, statistical sampling, computer simulation, and computer-based games \cite{RNG_uses}.  This is another area in which QCs are theoretically superior to CCs, as the probabilistic nature of quantum mechanics allows for a QC to operate as a Quantum Random Number Generator (QRNG). A QRNG would produce truly random numbers. 
\par
QRNGs have been widely investigated and discussed in scientific literature. A comprehensive review and classification of QRNGs was recently performed in \cite{QRNG_review}. Much work has been done to investigate the efficacy of optical-based QRNG apparatus and found promising results, such as in \cite{Bayesian_Criteria, optical_QRNG_2}. Much less work has been done to explore QRNGs in more complex systems, such as existing large-scale quantum computers.  
\par
In this work, We discuss the development of a QC efficacy testing method based on QRNGs and the application of the method to several of IBM's Quantum Computers. The rest of the paper will be structured as follows. In Section II, We discuss the theory behind the testing method, including the QRNG algorithms and statistical tests utilized. In Section III, We discuss the interpretation of the statistical tests and report their results. In Section IV, we discuss the results of the testing method and their implications.


\section{Theory and application of the Method}


\subsection{The Basic QRNG}

Many different QRNGs may be conceived, but the simplest one requires only a single qubit. At the start of the algorithm, the qubit is initialized in either the 1 or the 0 states (IBM's QCs initialize in the 0 state). Next, the qubit is operated on by a Hadamard Gate, whose function is to put the qubit into a superposition, such that each basis state of the qubit (a 1 or a 0) is equally likely to be the result of measurement. Then the qubit is operated on by a measurement gate, which makes a measurement of the state of the qubit, collapsing the superposition. The result of the measurement is then stored classically. In this work, we shall refer to this algorithm as the Basic QRNG (see Fig. \ref{fig:QRNG circuits} a). Repeating this algorithm results in a random binary sequence. This ability of QRNGs to be truly random is a direct consequence of the probabilistic nature of the measurement collapse principle. After the operation of the Hadamard Gate, the qubit is in the state. 

\begin{equation}
    | \Psi \rangle= \frac{1}{\sqrt{2}} |0\rangle + \frac{1}{\sqrt{2}} |1\rangle,
\end{equation}

where each state carries $\frac{1}{2}$ of the probability. In an ideal system, measurement of the qubit's state has exactly $50\%$ probability of resulting in either basis state. 


\subsection{Efficacy Testing Method}
However, any physical system, such as real QC, is non-ideal and introduces biases such that the probability is not exactly $50\%$ for each state \cite{Bayesian_Criteria}. A high-efficacy QC, if properly implemented and constructed, should be sufficiently devoid of defects and biases so that they have a negligible impact on the output of a QRNG. Whereas a low-efficacy QC could have significant biases which would cause a skew in its outputs. Therefore, one can investigate the efficacy of a QC by analyzing the randomness of QRNGs run on the QC to detect any bias or patterns. This is the premise of the QC efficacy testing method presented here.
\par
The method is accomplished by means of statistical tests, which are algorithms developed to examine a data sequence for certain properties, in particular, detectable statistical patterns\cite{Bayesian_Criteria, NIST, TBT}. In line with standard terminology from related works, we will use the following terms: "statistically random" (SR) and "statistically non-random" (SNR).  SR, in this context, means that no statistical patterns have been found in the data, whereas SNR indicates that a pattern or a predictable element has been observed. Each statistical test looks for a particular pattern or property and reports if the sequence passes and should be considered SR, or fails and should be considered SNR. The degree to which it passes or fails is also reported by the tests.
\par
Different statistical tests are developed from different formulations of probability theory. Even tests developed from similar theories look for different patterns and properties and may be considered more or less "strict" than other tests. Recognizing these variations, our method applies a battery of different statistical tests.  Additionally, given that a truly random source could produce a sequence that would fail statistical tests (i.e. a sequence of all ones is just as likely as any other sequence but would be found as SNR), our methods performs statistical tests on multiple QRNG outputs before a conclusion can be made about a particular source. 


\subsection{Additional QRNGs}
While the simplest QRNG is the Basic QRNG, other QRNG algorithms may be used to investigate certain abilities of QCs or come closer to true randomness. We outline three additional QRNGs that will be utilized in this work. 
\par
Preliminary results suggested that the QCs (running the Basic QRNG) are biased towards 0s in the output. We hypothesize that this bias is an artifact of the qubits being initialized in the 0 states and that the Hadamard Gate does an imperfect job of placing the qubit into an even 50-50 superposition from that state. This can be investigated by using the following algorithm in the efficacy testing method. 
\par
First, the qubit is initialized. Then it is operated on by a NOT Gate (whose function is to flip the qubit from a 1 to a 0 or from a 0 to a 1) before passing it to the Hadamard Gate, and then to the Measurement Gate. Now all the qubits will be in the 1 state when they are operated on by the Hadamard Gate. We will refer to this algorithm as QRNG Type 1 (see Fig. \ref{fig:QRNG circuits}b). If the bias towards 0s is due to the initialization of the qubit, then we should see a bias towards 1s from this algorithm. 
\par
Because the functionalities of a QC rely on superposition, it is natural to investigate various aspects of the performance of the Hadamard gate (as it is the mechanism by which superposition is obtained in a QC). One such aspect is the ability of a Hadamard gate to return a qubit to a superposition after it has already been measured. This can be done by utilizing the following algorithm. After initialization, the qubit is operated on by a Hadamard gate, then a Measurement gate (whose results are not stored), then another Hadamard gate, and then a final Measurement Gate (storing the result this time). We will refer to this algorithm as QRNG Type 2 (see Fig. \ref{fig:QRNG circuits}c). 
\par
The previous algorithms have utilized only a single qubit, whereas the QCs used for this study have a total of either 5 or 7 qubits (more on these QC systems in Section 3.). Utilizing more qubits may bring us closer to realizing true randomness. This can be investigated using a four-qubit algorithm where two of the qubits follow the process of the Basic QRNG and two of the qubits follow the process of QRNG Type 1 (see Fig. \ref{fig:QRNG circuits}d). We will refer to this algorithm as QRNG Type 3. This construction may counteract some of the previously observed bias, as well as create more data to enable a more thorough analysis (as now each execution of the algorithm returns four bits of data instead of one).


\subsection{Application of the Method}
We applied the aforementioned QC efficacy test to nine IBM Quantum systems, utilizing a battery of 18 statistical tests detailed in \cite{Bayesian_Criteria, NIST, TBT}. See the Table \ref{table:statistical test battery} for details on each statistical test. These IBM systems are different physical QCs, using 3 different processor architectures. Each "trial" represents one execution of the given QRNG, which repeats the circuit 8,192 times (the number of "shots" allowed at our access level) and returns a data sequence consisting of the results of each run, chronologically. This yields each trial as a binary sequence of 8,192 digits, or 32,768 digits in the case of QRNG Type 3. By analyzing a large number of trials, we can determine how SR the systems are, and therefore the efficacy of each system. See Table \ref{table: QC info} for the different QCs, their processors, the number of qubits, and the number of trials for each QRNG for that QC. Note that some QCs have more trials for a given QRNG, this is due to extra trials having been run before the decision to generally use 128 trials was made. 
\par
The battery of statistical tests was applied to each individual trial. Additionally, trials were strung together (to form sequences longer than 8,192 bits) and tested, to fit the minimum sequence lengths required by certain tests. We refer to these longer sequences as "combined," and are explained further in section III.A.
\par


\begin{table*}
\renewcommand{\arraystretch}{1.5}
\begin{center}
\begin{tabular}{||>{\centering\arraybackslash}p{.75 in} >{\centering\arraybackslash}p{.5 in} >{\centering\arraybackslash}p{1. in} >{\centering\arraybackslash}p{.75 in} >{\centering\arraybackslash}p{.75 in} >{\centering\arraybackslash}p{.75 in} >{\centering\arraybackslash}p{.75 in}||} 
 \hline
 QC & Qubits & Processor & \centering{No. of Trials for Basic QRNG} & \centering{No. of Trials for QRNG Type 1}& \centering{No. of Trials for QRNG Type 2}& No. of Trials for QRNG Type 3\\ [0.5ex] 
 \hline\hline
 
  Perth & 7 & Falcon r5.11H & 128 & 128 & 128 & 128\\ 
 \hline
  Lagos & 7 & Falcon r5.11H & 128 & 128 & 128& 128\\ 
 \hline
  Nairobi & 7 & Falcon r5.11H & 206 & 128 & 128& 128\\ 
 \hline
  Oslo & 7 & Falcon r5.11H & 256 & 128 & 128& 255\\ 
 \hline
  Jakarta & 7 & Falcon r5.11H & 128 & 128 & 128 & 128\\ 
 \hline
  Manila & 5 & Falcon r5.11L & 325 & 128 & 128& 131\\ 
 \hline
  Quito & 5 & Falcon r4T & 128 & 128 & 128& 128\\ 
 \hline
  Belem & 5 & Falcon r4T & 128 & 128 & 128& 128\\ 
 \hline
  Lima & 5 & Falcon r4T & 319 & 128 & 128 & 133\\ 
 \hline
\end{tabular}
\caption{\small{ The QCs utilized in this work and supplemental information about them and the number of trials used.}}.
\label{table: QC info}
\end{center}
\end{table*}

\begin{figure*}
    \centering
    \includegraphics[scale=.37]{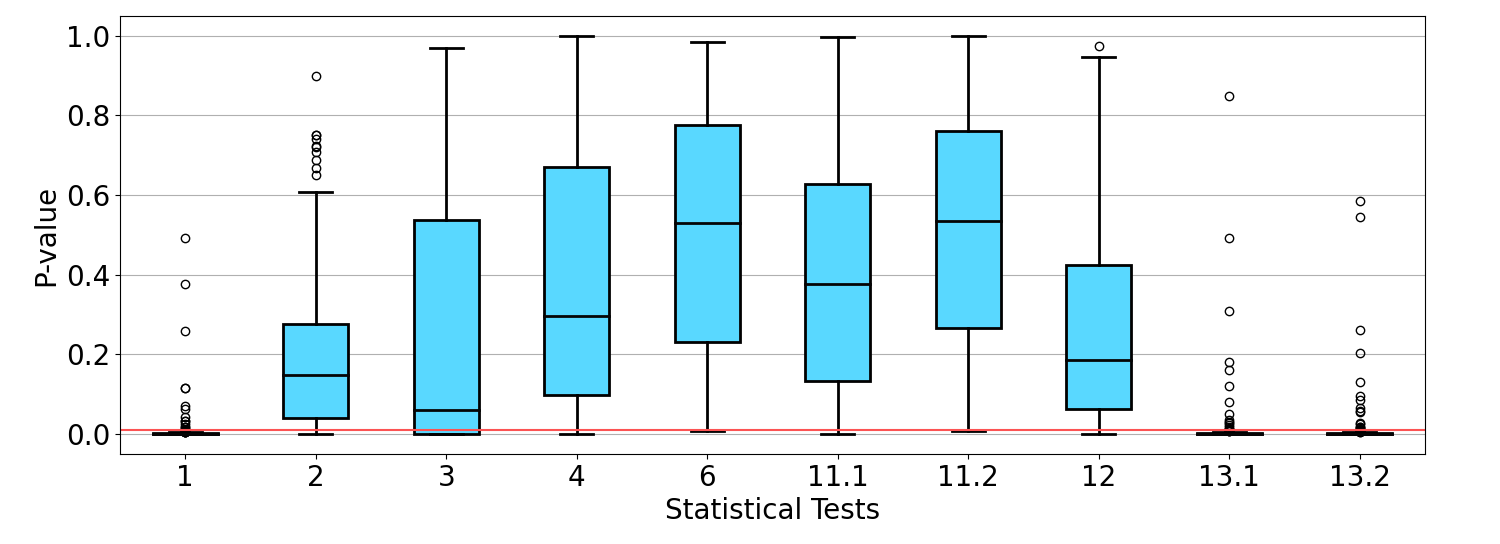}
    \includegraphics[scale=.41]{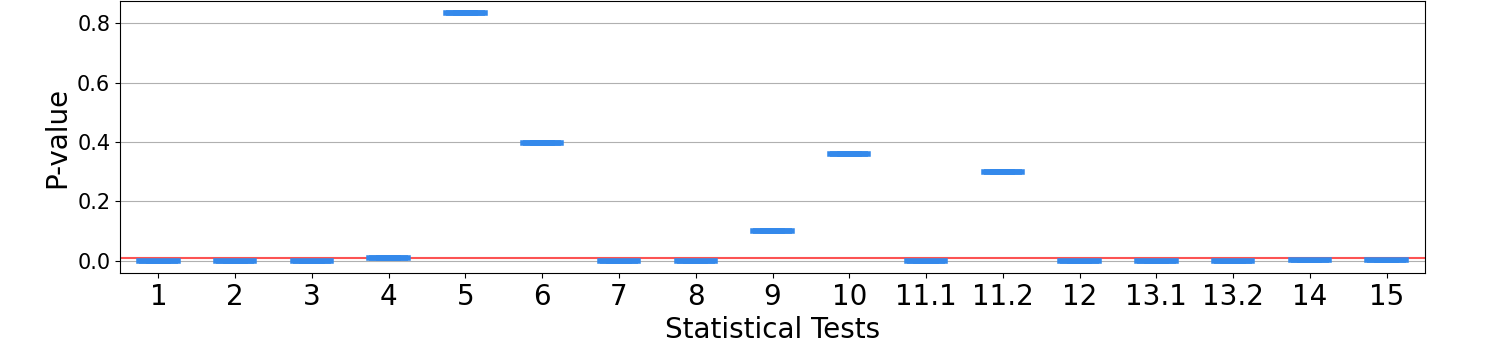}
    \caption{Results from tests 1-15 for the Basic QRNG from the Perth IBM QC system at the trial (top) and combined (bottom) levels. The box plots represent the results of individual trials for each statistical test. The horizontal axis labels for the statistical tests reference Table \ref{table:statistical test battery}. The red line is at a P-value of 0.01.}
    \label{fig:NIST Perth Basic}
\end{figure*}

\begin{figure*}
    \centering
    \includegraphics[scale= 0.5 ]{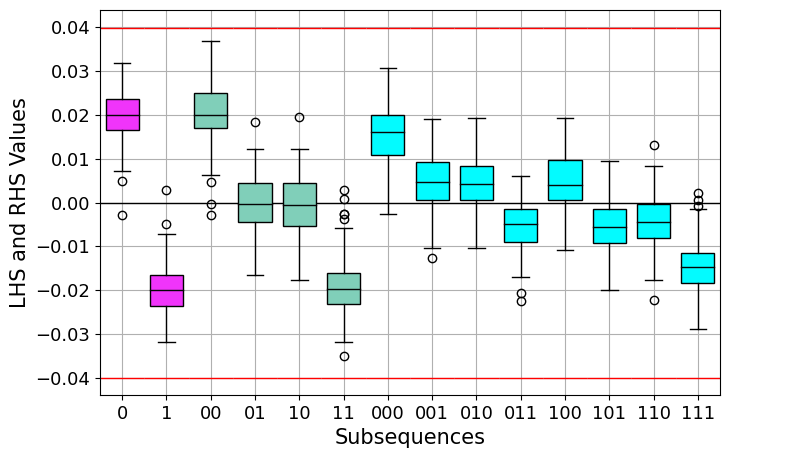}
    \includegraphics[scale=0.65]{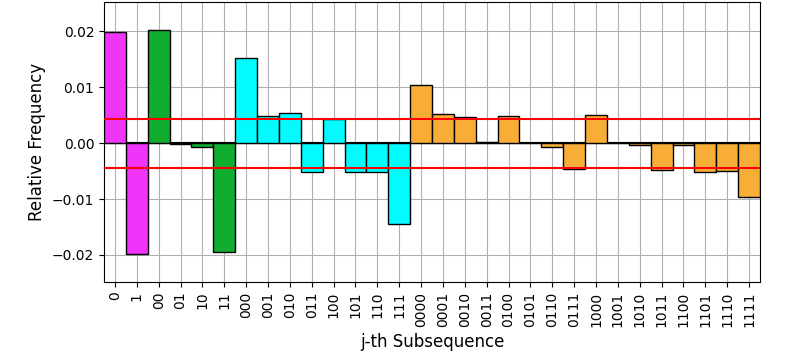}
    \caption{Results from test 16 for QRNG Type 1 from the Jakarta IBM QC System at the trial level (top) and the combined level (bottom). The bars represent the LHS values for each subsequence and are color-coded by each subsequence length. The RHS bound is represented by the red lines.}
    \label{fig:Bor Jakarta Type 1}
\end{figure*}


\section{Results}


\subsection{Interpretation and Usage of Statistical tests}

Many times in this section, the results from a multitude of trials are presented together. For these cases, a system is considered to have failed a statistical test when the median result for its trials would be found as SNR. The results have been summarized in Table \ref{table:results table}. In this section, we consider the interpretation of the statistical tests that led the summarized results. 
\par
A detailed example of the results of statistical tests 1-15, from \cite{NIST}, is given in Figure \ref{fig:NIST Perth Basic} for the Basic QRNG from Perth QC. Note that the statistical tests included in the top of this Figure \ref{fig:NIST Perth Basic} are the ones that function for sequence lengths of 8,192, or 32,768 digits for QRNG Type 3. Here each test reports a "P-value" that indicates how SR a test is. The "decision point" is a P-value of 0.01; scoring below this threshold represents a failure and indicates the sequence as SNR. The P-value is a statement about how likely the sequence was to be generated randomly.
\par
There are several items to note regarding tests 1-15. First, tests 7, 14, and 15 return multiple P-values (see \cite{NIST} for further details). We have chosen to include the strict interpretation of these P-values, meaning if at least one of the P-values from these tests is below the decision point, then the sequence is found to be SNR by the given test. The lowest P-value for tests 7, 14, and 15 is shown in the figures in this work.
\par
Second, tests 2, 10, 11, and 12 operate by examining all the sections of the sequence that are of a certain size, which is an input parameter for each test. Here we write the parameter generally as $M$, despite there being a different parameter for each of these tests. For example, if $M=5,000$ for test 10, the Linear Complexity Test, the test would look at each section of up to 5,000 digits in the sequence. Each of these tests has a range of allowed values for $M$ (specified in their source material documentation) which the test is valid for. Using larger values of $M$ increases the strictness of each test. As before, We have chosen to include the strict interpretation for these tests and utilize the largest allowed $M$ values. 


\begin{figure*}
    \includegraphics[scale=0.52]{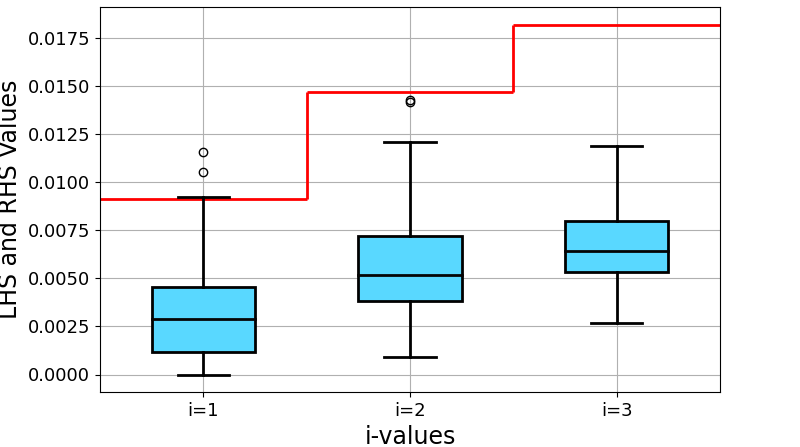}
    \includegraphics[scale=0.45]{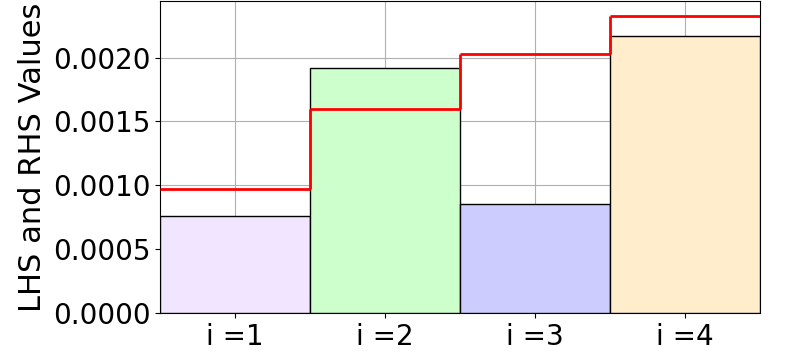}
    \caption{Results of test 17 for the combined QRNG Type 3 trials for the Lagos IBM QC system at the trial level (top) and the combined level (bottom). The bars represent the LHS values for each \textit{i}-value. The RHS values are represented by the red line.}
    \label{fig:Bay Lagos Type 3}
\end{figure*}


A detailed example of the results of test 16, the Borel Normality Criteria from \cite{Bayesian_Criteria}, are reported in Figure \ref{fig:Bor Jakarta Type 1} for the Type 1 QRNG from the Jakarta QC. The Borel Normality Criteria functions by calculating a value based on the frequency for each possible \textit{i}-lengthed subsequence. Only \textit{i}-values up to a certain limit are required, based on how long the sequence is, with longer sequences requiring the use of larger \textit{i} values.  We refer to this first value as the "LHS value." The LHS value is compared with the "RHS value," calculated from expectations about the subsequence's frequency for a random case and the length of the sequence. A larger \textit{positive} LHS value indicates an overabundance of the respective subsequence and corresponds to a larger \textit{negative} LHS value for another subsequence (which appears less). If the LHS value is within the LHS value bounds, the sequence is considered to be SR. 
\par
A detailed example of the results of test 17, the Bayesian Criteria from \cite{Bayesian_Criteria}, are shown in Figure \ref{fig:Bay Lagos Type 3} for the Type 3 QRNG from the Lagos QC. The Bayesian Criteria test operates similarly to the Borel Normality Criteria except that instead of considering the frequency of individual subsequences for its LHS value, it considers the relative frequency of \textit{all} subsequences of length \textit{i} (again, up to a certain appropriate length). It also calculates a different RHS value for each \textit{i}-value. If the LHS value is less than the RHS value, the sequence is considered SR. The Bayesian Criteria is considered stricter than the Borel Normality Criteria.
\par
\begin{figure*}
    \centering
    \newcommand{\myscaleIII}{0.5}
    \includegraphics[scale=\myscaleIII]{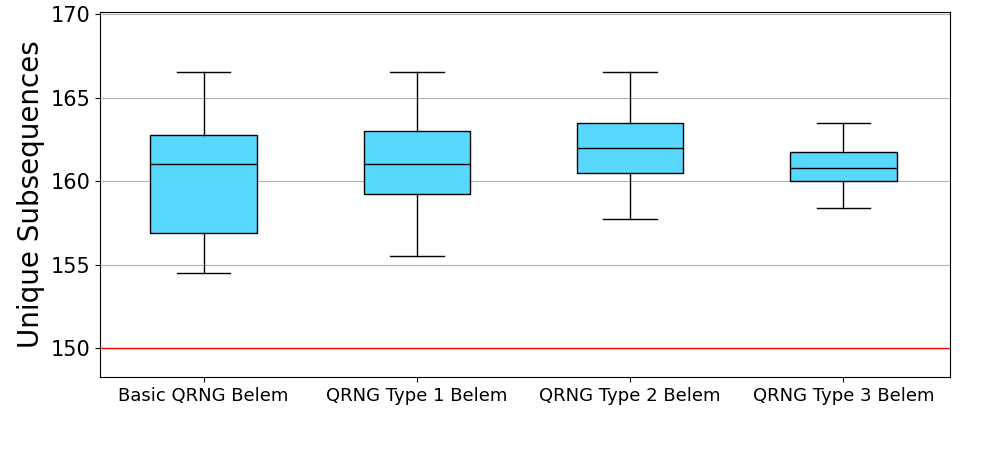}
    \renewcommand{\arraystretch}{0.8} 
    \begin{tabular}{||>{\centering\arraybackslash}p{.5 in} >{\centering\arraybackslash}p{.75in} >{\centering\arraybackslash}p{1.25in} >{\centering\arraybackslash}p{1 in}||}
    \hline
    \rule{0pt}{3ex} System & QRNG & Unique Sub-sequences & SR or SNR? \\  
    \hline\hline
    \rule{0pt}{3ex} Belem   & Basic & 36018 &  SNR \\ 
         \rule{0pt}{3ex}   & 1     & 41155 &  SNR \\ 
         \rule{0pt}{3ex}   & 2     & 41189 &  SNR \\ 
         \rule{0pt}{3ex}   & 3     & 41040.25 &  SNR \\ 
    \hline
    \end{tabular}

    \caption{Results of test 18 for all of the QRNGs from the Belem QC at the trial level (top) and the combined level (bottom). For the trial level, the box plots represent the results of individual trials for each statistical test. The red line represents the threshold of 150 unique subsequences. }
    \label{fig:TBT Belem }
\end{figure*}

A detailed example for the results of test 18, the Topological Binary Test (TBT) from \cite{TBT}, are reported in Figures \ref{fig:TBT Belem } for the Belem QC. TBT operates by comparing the number of unique subsequences of a certain length \textit{m} (determined by the length of the sequence) and comparing it to a certain critical value. This threshold is found from an expectation of how many unique subsequences would be present in a randomly generated sequence of the same length. If there are more unique subsequences than the critical value, the sequence is considered SR. Note that the TBT test only operates on sequences of exactly certain lengths for different values of $m$ (i.e. for $m=8$ the sequence must be 2,048 digits, or for $m=15$ the sequence must be 491,520 digits). 
\par
For the case of trials that are 8,192 digits long, each trial was partitioned into four sequences of 2,048 digits, or for QRNG Type 3 trials (which are 32,768 digits long) 16 sequences of 2,048 digits. This works because 8,192 happens to be an exact multiple of 2,048, which, as far as the authors are aware, is purely coincidental (none of the other sequence lengths allowed by the TBT test are exact factors of the lengths of the trials). These partitioned sequences were then tested by the TBT test using \textit{m}=8. The number of unique subsequences for each of the partitioned sequences was then averaged and compared to the \textit{m}=8 critical value of 150 unique subsequences. This was done to most efficiently utilize the information in all of the trials. 
\par
As mentioned earlier, in addition to analyzing all single trials from a QC-QRNG combination via statistical tests, longer "combined" sequences were constructed by stringing together (chronologically) all the trials from each of the QC-QRNG pairings. These combined sequences allow for the use of statistical tests that require longer sequences to function. Additionally, some tests become more effective at diagnosing statistical non-randomness when applied to longer sequences. These are also shown in Figures \ref{fig:NIST Perth Basic} - \ref{fig:TBT Belem } for each of the statistical tests, respectively.
\par
In regards to how to interpret the results from the statistical tests, there are several things that I will note. Because the tests tell us how likely it is that a sequence was generated randomly, failing a small portion of the tests does not necessarily mean that the sequence was not generated randomly. In particular, failing only one or two of tests 2, 7, 10, 11, 12, 14, and 15 (as we are representing the strict interpretation of these tests) does not necessarily exclude a sequence from being considered SR. It is only if a sequence fails many tests, or if many sequences from a given source repeatedly fail a test, that we can conclude them to be SNR. Ideally, a truly random sequence will pass all the statistical tests, but given that a truly random source is just as likely to generate a sequence that would fail the tests (say a sequence of all 1s), then we can only conclude that a source is SR after observing that it is rare for it to fail a test. 
\par
The results of all the statistical tests for every QC-QRNG combination have been compiled in Table 2. Note that Table 2 only shows whether a combination passed or not and does not include information on "how much" a combination passed or failed by. 


\begin{table*}
    \includegraphics[scale=.95]{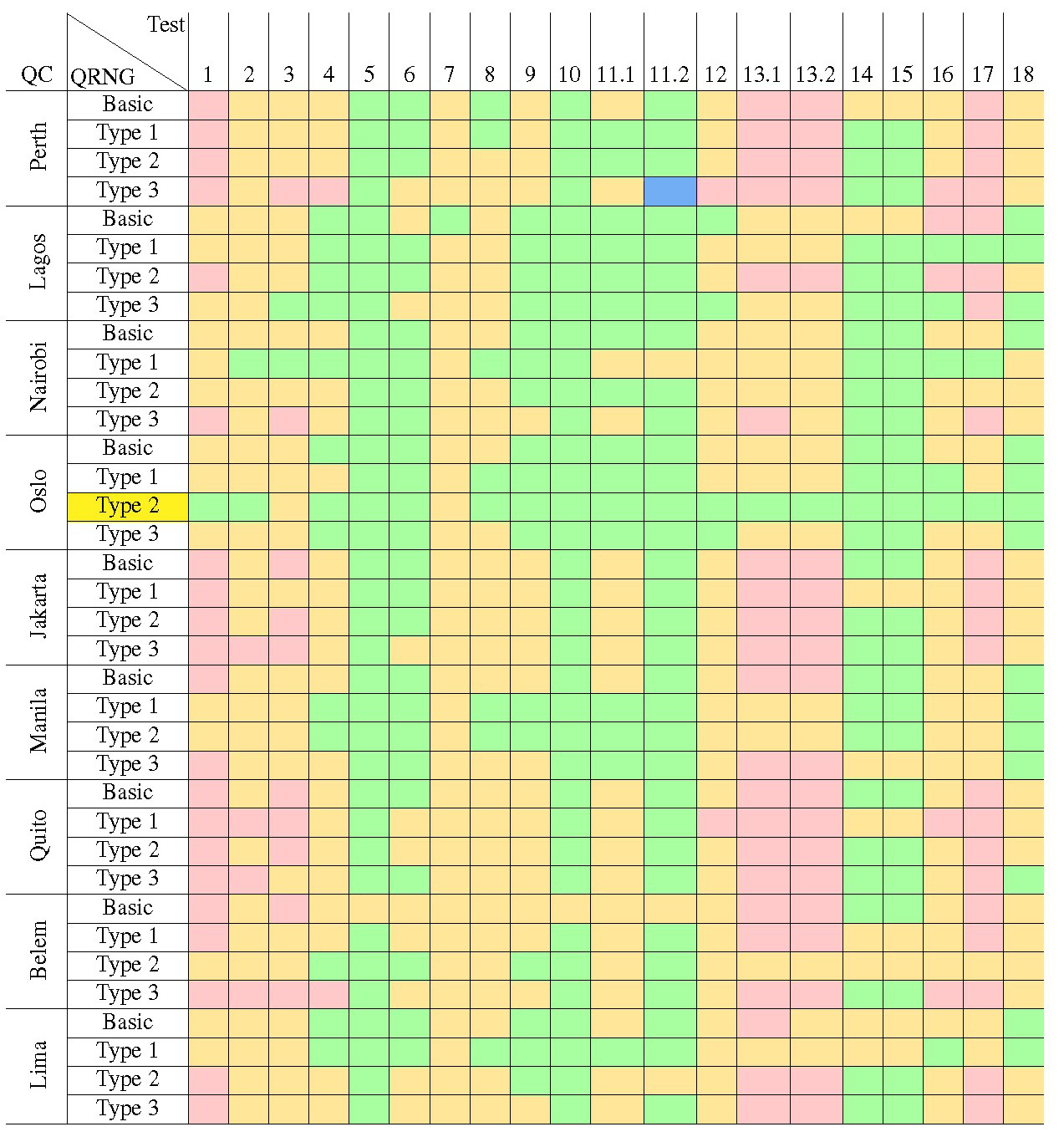}
    \caption{The results of the statistical test battery for each of the QC-QRNG pairings. The color coding indicates as follows: red, failed a given test at both the trial and combined sequence levels; yellow, passed at the trial level but failed at the combined; green, passed at both levels; blue, failed at the trial level but passed at the combined. Tests 5, 7-10, 14, and 15 only operate at the combined level and therefore are only yellow and green if the QC-QRNG failed or passed, respectively. Oslo Type 2 has been highlighted for its significance.}
    \label{table:results table}
\end{table*}

\begin{table*}
    \includegraphics[scale=.95]{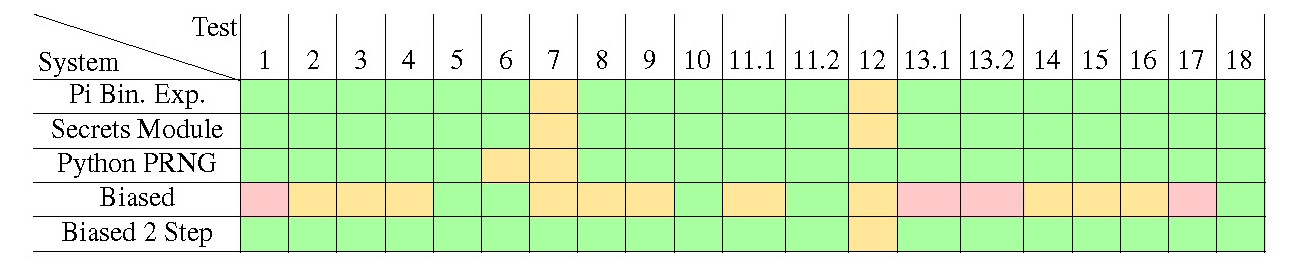}
    \caption{The results of the statistical test battery for each of the Comparison systems. Again, the color coding indicates as follows: red, failed a given test at both the trial and combined sequence levels; yellow, passed at the trial level but failed at the combined; green, passed at both levels; magenta, failed at the trial level but passed at the combined. Tests 5, 7-10, 14, and 15 only operate at the combined level and therefore are only yellow and green if the comparison system failed or passed, respectively.}
    \label{table:comparison results table}
\end{table*}


\subsection{QC Efficay Rankings}
It can be seen that the different QC systems demonstrate varying levels of statistical randomness, which depends further upon the QRNG algorithm. The Basic QRNG yields the most SR outputs for Oslo and Lagos. QRNG Type 1 yields the most SR outputs from Lagos, Nairobi, Oslo, and Lima. QRNG Type 2 yields the most random outputs from Oslo. QRNG Type 3 yields the most SR outputs from Lagos and Oslo (and yields highly SNR outputs for all the other systems). Despite this variation, we can still generally group the systems by efficacy; with Lagos and Oslo being of high efficacy; Lima, Nairobi, Manila, and Belem being of intermediate efficacy; and Quito, Perth, and Jakarta being of low efficacy. Only a single QC-QRNG combination is found as SR, QRNG Type 2 from Oslo. Additionally, Oslo is the only QC that always passes at the trial level.


\section{Discussion}


\subsection{Ranking and Performance of the QCs}

While a truly exhaustive discussion of the implications of the results of the statistical tests for each QC - QRNG combination is a significant task, suitable for its own work, here we summarise several key conclusions. Some of the higher efficacy systems perform very well at the trial level for some of the QRNGs, some well enough that it suggests the QC could be capable of true randomness. However, when considering the combined trial sequences for each QC - QRNG pairings, all but one are overwhelmingly found to be SNR. The only one that is found to be SR is QRNG Type 2 from Oslo.
\par
QRNG Type 2 from Oslo passes every test at the trial level and only fails two tests at the combined level, which is not enough to conclude it to be SNR. The tests that it fails are tests 3 and 7. This is momentous and suggests that QRNG Type 2 ran on the Oslo QC may indeed be a realization of a truly random physical system. Further testing should be done to determine whether this result is replicable.


\subsection{Comparison Systems}

In order to provide a point of comparison for the QCs, We applied the statistical tests to a series of systems, which we refer to as the "comparison systems". The first comparison system is the binary expansion of the digits of pi. Given that there are no known patterns among the digits of pi, it is used as a "gold standard" of randomness \cite{NIST}. It is important to note, that while we take the decimal digits of pi to be statistically random, the conversion from decimal digits to binary results in repetitions of subsequences which are the binary forms of respective decimal digits. This causes the binary expansion of pi to be found as SNR by test 12, despite it being SR in nature. 
\par
The second comparison system is Python's built-in PRNG, which operates on the Mersenne Twister Algorithm and is known to be difficult to show as SNR, but not SR enough to be considered cryptographically secure \cite{python_docs_random}. This system demonstrates the performance of a high-quality, but not state-of-the-art, PRNG. 
\par
The third comparison system is the PRNG found in Python's Secrets module, which utilizes a computer's low-level operating system security protocols to operate PRNGs (so on a Windows computer, as was used here, the module uses Window's cryptographically secure PRNG) \cite{python_docs_secrets}. This system illustrates the performance of a PRNG of the highest class. 
\par
The fourth comparison system is an intentionally SNR PRNG, generated to have a bias of 52\% towards 0s and 48\% against 1s. This was done utilizing the Python module Numpy \cite{numpy} and will be referred to as "PRNG Biased." This system serves as an example of the results that would be seen from a non-random system. 
\par
The fifth comparison system is similar in nature to PRNG Biased but adds a second step. First, a "precursor" bit is generated as 1 or 0 with a 0.52 bias towards 0 and 0.48 bias against 1s. Then, if the precursor bit is a 0, the "final" bit is generated with the same bias as the precursor bit, but if the precursor bit is a 1, then the final bit is generated with a 0.48 bias away from 0s and 0.52 towards 1s. The final bit is then included in the binary sequence. We will refer to this system as "PRNG Biased 2 Step." This system illustrates what happens when successive non-random steps counteract each other. 
\par
The data from each comparison system was generated in 128 trials of 8,192 digits. These trials were individually analyzed by the statistical tests, as well as in a combined form. This was done to mimic the format of the data from the QCs. The results of the statistical test battery applied to these comparison systems are shown in Table 3.


\begin{figure*}
    \includegraphics[scale=0.41]{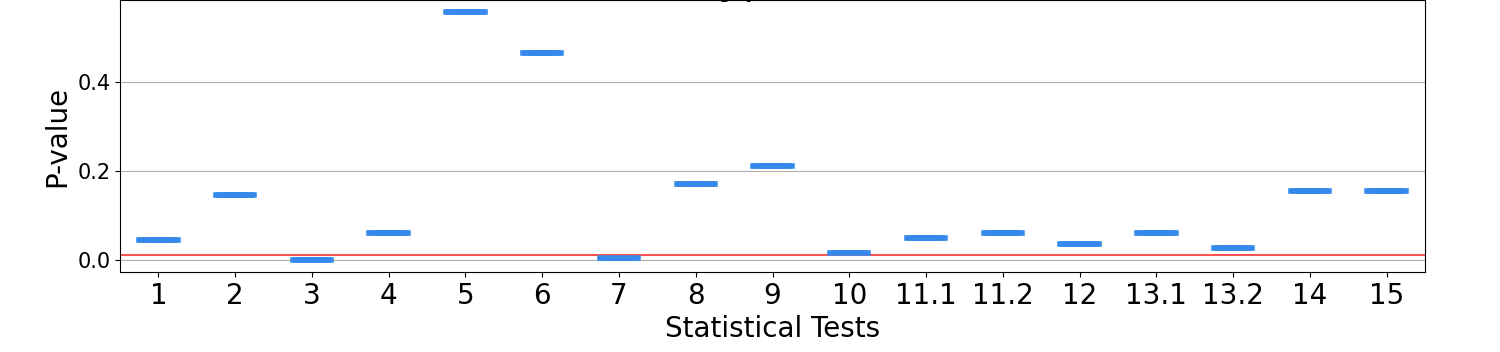}
    \includegraphics[scale=0.41]{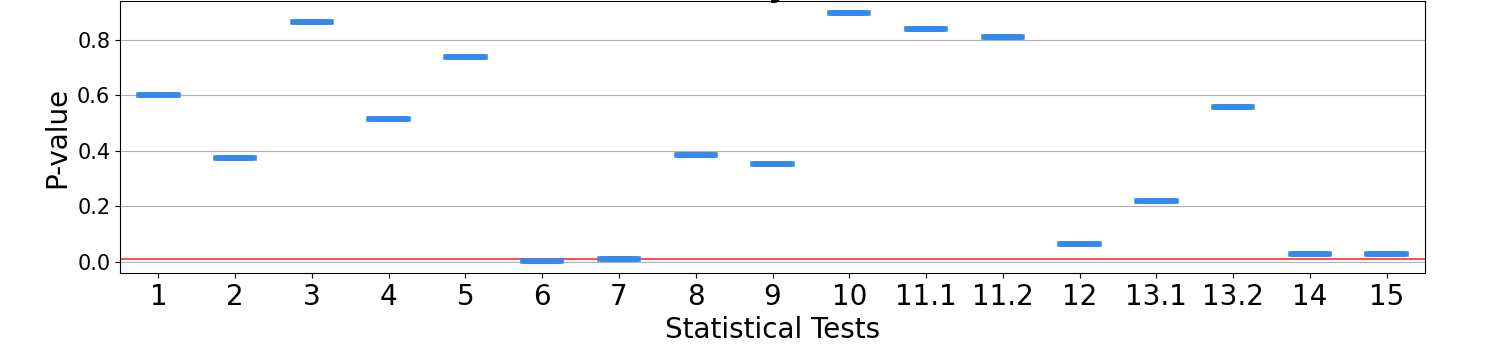}
    \caption{Results from tests 1-15 of the QRNG Type 2 from the Oslo IBM QC system (top) and Python's PRNG (bottom) at the combined level. The box plots represent the results of individual trials for each statistical test. The indexes for the statistical tests reference Table 1. The red line is at a P-value of 0.01. Notice that while both systems pass all but 2 tests, Python's PRNG tends to have higher P-values }
    \label{fig: Oslo Type 2 and Pi}
\end{figure*}


As expected, the binary expansion of pi, the Secrets module, and Python's PRNG all perform very well, passing all the tests at the trial level. At the combined level, Pi and the Secrets module fail tests 7 and 12, but, as discussed earlier, because of the strict interpretation of these tests used in this work, this result is not unusual. Python's PRNG fails tests 6 and 7 at the combined level, which is consistent with it being less random than the previous two. PRNG Biased fails four tests at the trial level, and nearly all of them at the combined level. Clearly, even 2\% deviation from an evenly split probability of bits being 1s or 0s is easily detected by the statistical tests. 
\par
Curiously, PRNG Biased 2 Step is found to be more SR than the other comparison systems, failing only a single test, test 12, at the combined level, even though we know it to have been generated by a nonrandom algorithm. This demonstrates an interesting phenomenon: successive non-random processes can effectively "cancel out" each other's non-randomness, and result in statistically random outputs. We refer to this phenomenon as "bias counteraction." 


\subsection{Inferences about QCs}

By analyzing the specific ways that the QCs are found to be SR or SNR, and by comparing them with the comparison systems, several things are inferred about the QCs
\par
Inference 1. The fact that many QC-QRNG pairings perform well at the trial level, but very poorly at the combined level suggests that whatever the form of the bias in the outputs is, it is not impactful within each trial. That is, the bias does not create any patterns within a given trial, but the bias repeats itself for each trial.
\par
Inference 2. The comparison systems (excluding PRNG Biased), were all found to be SR to a high degree, despite the fact that (aside from Pi) it is known that they were generated from non-random sources. QRNG Type 2 from Oslo, despite being the only system to be found as SR, did not display as high a degree of randomness as the comparison systems. This can be seen in Figure \ref{fig: Oslo Type 2 and Pi} and demonstrates that, at least for these QCs, operating these QRNGs, "quantum supremacy" has yet to be realized for the purpose of random number generation.  
\par
Inference 3. Every QC-QRNG combination except the Basic QRNG and QRNG Type 2 from Oslo, and the Basic QRNG and QRNG Types 2 and 3 from Lagos shows an overabundance of 0s (which can be seen from the test 16 LHS values at the combined level). This suggests that for the rest of the QCs, there is a significant source of bias towards 0s. This could potentially be caused by either the Hadamard gate being faulty and creating a superposition state with a bias towards 0s, the presence of noise is the system manifesting as 0s, both, or something else. Notice that the results from the PRNG Biased system closely resemble the results for many of the QC-QRNG pairings. This provides further evidence that the bias for many of the systems manifests as each qubit measurement having a bias towards 0s. 
\par
Inference 4. Several QCs systems, Nairobi, Manila, and Lima, show an improvement in performance from QRNG Type 1 over the Basic QRNG. This suggests that, for these systems, the bias towards 0s does in fact come from the Hadamard gate creating a superposition state with a bias toward the initial state. Furthermore, several systems, Nairobi, Oslo, Lagos, Lima, and Belem, all exhibit more random results from QRNG Type 2 than Type 1 or the Basic QRNG, while the other systems perform similarly for Type 1 and 2. These systems which perform better with QRNG Type 2 may be due to the effect of bias counteraction, resulting from the successive Hadamard gates, similar to what is observed in the Biased 2 Step comparison system.
\par
It should be noted that while the effect of bias counteraction that is observed in QRNG Type 2 and PRNG Biased 2 Step is similar, the underlying mechanism is different. In PRNG Biased 2 Step, the bias counteraction works by making the system more chaotic, obfuscating the fundamentally deterministic nature of the algorithm, and making it appear SR. However, in the case of QRNG Type 2, the bias counteraction works to "filter out" the noise present in the system, or whatever the source of the bias may be, distilling the system to its fundamentally statistical nature.
\par
Inference 5. Almost every QC is found as heavily SNR for QRNG Type 3 with only Lagos and Oslo performing reasonably well at either the trial or combined level. Considering that several systems, Lagos, Lima, Oslo, and Perth, fail Tests 16 and 17 at the $i=4$ by a much larger margin than any other $i$ levels, this indicates that there are repetitions of 4 digit patterns in the resulting data. This suggests that the state of the qubits is not sufficiently independent of each other and that the qubits are becoming unintentionally entangled or otherwise correlated to some degree. For the other systems, either this entanglement is not occurring, or perhaps, considering that they perform worse for QRNG Type 3 than the other QRNGs, the source of the bias is significant enough to prevent the 4-digit patterns from being apparent. Another possibility is that the bias increases with the number of qubits in use. 


\subsection{Conclusion}

In this work, we proposed a method of investigating the efficacy of QCs, by utilizing their theoretical potential to demonstrate true randomness. The method involved repeatedly executing a QRNG on the computer and analyzing the statistical randomness of its results, with the analysis being done using a battery of 18 statistical tests. 
\par 
We then applied this efficacy testing method to nine IBM quantum computers, Perth, Lagos, Nairobi, Oslo, Jakarta, Manila, Quito, Belem, and Lima, using four different QRNG algorithms, each designed to investigate certain properties of the QCs. All but one of the QC-QRNG pairings were found to be statistically non-random, with the output from the QRNG Type 2 algorithm ran on the Oslo QC being found as statistically random. This suggests that this QC-QRNG combination may indeed be a physical realization of true randomness. 
\par
We compared the statistical randomness of the QCs to five comparison systems, the binary expansion of pi, a PRNG using Python's Secrets module, a PRNG using Python's built-in PRNG, a biased "PRNG," and a PRNG whose biases cancel out. None of the quantum computers performed better than these sources (aside from the biased PRNG), QRNG Type 2 from Oslo being found as random and coming close in terms of statistical randomness. This indicates that further work is necessary in order for these quantum computers to fully realize their theoretical potential to be truly random, either by addressing biases present in the hardware of the QCs or by utilizing quantum algorithms that are capable of extracting the true randomness through the bias. QRNGs utilizing repeated Hadamard gates, such as QRNG Type 2, could be a promising direction for achieving such algorithms.
\par

\section*{Acknowledgements}
We acknowledge the use of IBM Quantum services for this work. The views expressed are those of the authors, and do not reflect the official policy or position of IBM or the IBM Quantum team.\\[13cm]

\begin{table}[H]
\footnotesize{
\begin{tabular}{p{0.25\linewidth} | p{0.65\linewidth}}
 \hline
 \textbf{Statistical Test} & \textbf{Function} \\ 
 \hline
 1. Monobit Test & Tests for even distribution of 1’s and 0’s \cite{NIST}\\
  \hline
 2. Frequency Test Within a Block & Tests for even distribution of 1’s and 0’s within subsequences\cite{NIST} \\
\hline
 3. Runs Test & Tests for uninterrupted sequences of 1’s or 0’s \cite{NIST} \\
 \hline
 4. Test for the Longest Run of Ones in a Block & Tests for the longest run of ones in a block of given sizes matches expectations of a random case \cite{NIST}\\
 \hline
 5. Binary Matrix Test & Tests the rank property of matrices made from subsequences. \cite{NIST} \\
 \hline
 6. Discrete Fourier Transform Test & Analyzes the peak heights in the Discrete Fourier Transform of the sequence. \\
 \hline
 7. Non-overlapping Template Matching Test & Tests for the number of occurrences of pre-specified target strings within non-overlapping blocks of the sequence \cite{NIST}\\
 \hline
 8. Overlapping Template Matching Test & Test for the number of occurrences of pre-specified target strings within overlapping blocks of the sequence \cite{NIST}\\
 \hline
 9. Maurer's "Universal Statistical" Test & Tests for the number of bits between matching patterns. \\
 \hline
 10. Linear Complexity Test & Tests for the length of a linear feedback shift register. \\
 \hline
 11. Serial Test & Tests for the frequency of all possible overlapping patterns of a specified length across the entire sequence \cite{NIST}\\
 \hline
 12. Approximate Entropy Test & Tests for the frequency of overlapping blocks of two adjacent lengths against the expected result for a random sequence \cite{NIST}\\
\hline
 13 Cumulative Sums Test & Tests for the maximal excursion (from zero) of the random walk defined by the cumulative sum of adjusted digits in the sequence \cite{NIST}. 13.1 and 13.2 designate the forward and backward versions, respectively.\\
 \hline
 14. Random Excursions Test & Tests for the number of cycles having exactly \textit{K} visits in a cumulative sum random walk \cite{NIST}\\
 \hline
 15. Random Excursions Variant Test & Tests for the total number of times that a particular state is visited in a
cumulative sum random walk\cite{NIST}\\
\hline
 16. Borel Normality Criterion & Tests for the distribution of individual subsequences.\cite{Bayesian_Criteria}. \\
 \hline
 17. Bayesian Criteria & Tests for the relative distribution of all subsequences (up to a certain length). \cite{Bayesian_Criteria} \\
 \hline
 18. Topological Binary Test & Tests whether the frequency of unique subsequences is higher enough.\cite{TBT} \\
 \hline
\end{tabular}}
\caption{The statistical test battery and respective descriptions.}
\label{table:statistical test battery}
\end{table}

\bibliographystyle{ieeetr}

\begin{thebibliography}{10}
















\bibitem{shor's} P. W. Shor, "Algorithms for quantum computation: Discrete logarithms and factoring," in \textit{Proceedings 35th Annual Symposium on Foundations of Computer Science}, IEEE Comput. Soc. Press, 1994, pp. 124–134.

\bibitem{google_shor} F. Arute, K. Arya, R. Babbush, et al., "Quantum supremacy using a programmable superconducting processor," \textit{Nature}, vol. 574, pp. 505–510, 2019. [Online]. Available: https://doi.org/10.1038/s41586-019-1666-5

\bibitem{quantum_curious} C. Hughes, J. Isaacson, A. Perry, R. F. Sun, and J. Turner, \textit{Quantum Computing for the Quantum Curious}, Springer, 2021.

\bibitem{IBM_Quantum} IBM Quantum. [Online]. Available: https://quantum-computing.ibm.com, accessed 16 October 2022.

\bibitem{quantum_educators} "Quantum for Educators," IBM. [Online]. Available: https://www.ibm.com/quantum/educators, accessed 16 October 2022.

\bibitem{RNG_uses} M. Mulko, "Random number generation: What are its functions and the fields of usage?," \textit{Interesting Engineering}, 13 May 2022. [Online]. Available: https://interestingengineering.com/innovation/random-number-generation, accessed 15 October 2022.

\bibitem{QRNG_review} V. Mannalath, S. Mishra, and A. Pathak, "A comprehensive review of quantum random number generators: Concepts, classification and the origin of randomness," \textit{arXiv preprint arXiv:2203.00261}, 2022.


\bibitem{Bayesian_Criteria} A. C. Martinez, A. Solis, R. Diaz Hernandez Rojas, A. B. U’Ren, J. G. Hirsch, and I. Perez Castillo, "Advanced statistical testing of quantum random number generators," \textit{Entropy}, vol. 20, p. 886, 2018. [Online]. Available: https://doi.org/10.3390/e2011088

\bibitem{optical_QRNG_2} M. Eaton, A. Hossameldin, R. J. Birrittella, et al., "Resolution of 100 photons and quantum generation of unbiased random numbers," \textit{Nat. Photon.}, vol. 17, pp. 106–111, 2023. [Online]. Available: https://doi.org/10.1038/s41566-022-01105-9

\bibitem{NIST} A. Rukhin, J. Soto, J. Nechvatal, M. Smid, and E. Barker, "A statistical test suite for random and pseudorandom number generators for cryptographic applications," \textit{Technical report}, Booz Allen and Hamilton Inc Mclean Va, 2001.

\bibitem{TBT} P. M. Alcover, A. Guillamon, and M. d. C. A. Ruiz, "New randomness test for bit sequences," \textit{Informatica}, January 2013.

\bibitem{python_docs_random} The Python Foundation, "random — Generate pseudo-random," \textit{Python Documentation}. [Online]. Available: https://docs.python.org/3/library/random.html, accessed 9 December 2022.

\bibitem{python_docs_secrets} Python Documentation, "secrets — Generate secure random numbers for managing secrets". [Online]. Available: https://docs.python.org/3/library/secrets.html, accessed 9 December 2022

\bibitem{numpy} Numpy Documentation. [Online]. Available: https://numpy.org/doc/stable/, accessed 9 December 2022


\end{thebibliography}

\end{document}